\documentclass[doublecol]{epl2} 
\usepackage{geometry} 
\usepackage{graphicx,color}
\usepackage{graphicx}
\usepackage{amssymb,amsmath}
\usepackage{float}
\usepackage{longtable}
\usepackage{braket}
\usepackage{gensymb}
\usepackage{comment}

\title{Cooling jug physics}
\shorttitle{Physics of cooling jug}
\author{ O. Luniachek\inst{1}, R. Timchenko\inst{1}, O. Golubov\inst{2,1}}
\institute{                    
	\inst{1} School of Physics and Technology, V. N. Karazin Kharkiv National University -- Svobody Sq. 4, 61022, Kharkiv, Ukraine\\
	\inst{2} Department of Aerospace Engineering Sciences, University of Colorado Boulder -- 429 UCB, Boulder, CO, 80309, USA
}

\pacs{nn.mm.xx}{First pacs description}

\abstract{We discuss the physics of the pot-in-pot cooler. By balancing temperature decrease due to evaporation and temperature increase due to heat exchange, we find the equilibrium temperature of the pot. In this simplified model, the cooling jug acts as a psychrometer, and the theoretical prediction of our model is in a good agreement with psychrometric tables. Next, we study dynamics of the jug cooling. The cooling rate is limited by water vapour diffusion through air, heat conduction through air, and heat conduction through the body of the jug. The derived rate of temperature decrease is in general agreement with the result of our experiment. In the end, we discuss some additional factors, such as capillary effects in the raw clay, water viscosity in the capillaries, and impact of complex shape of the jug.}

\begin{document}
	\maketitle
	\label{firstpage}
	
	\section{Introduction}
Keeping food cool has always been an effective way of extending its storage life, and different devices have been used for this sake throughout history.
Most of them consume energy to compensate for the energy exchange of the food with the environment owing to the zeroth law of thermodynamics, so that currently the energy spent for food preservation is about 0.5 kWh per person per day \cite{energy_we_consume}. 

Evaporative cooling constitutes one of the most effective and widely spread ways of lowering the temperature, with its manifestations ranging from cooling hot food by blowing on it to lowering our own body's temperatures by sweating, and from evaporating fluorocarbon refrigerants for retaining low temperature in common fridges to attaining the coldest temperature ever reached in ultra cold atom experiments \cite{cold_atom_experiments}.
	
	In this article we focus on the pot-in-pot cooler, which harnesses evaporative cooling by water to keep foods fresh with zero energy consumption. It is a system made of two clay pots and a wet sand as a separator between them. The jug cools down due to the evaporative cooling of water from pores on the surface of the outer jug, to which the water gets from the wet sand.
	Evaporation is caused by heating of the outer surface by heat fluxes from both the atmosphere and the inner pot.
	At first we explain working principle of such a system in the static case, i.e. when the temperature of the jug surface and the external air are in equilibrium. In this model we consider only the contribution of heat flow from the atmosphere to water evaporation. Then we make our model dynamic by taking into account heat fluxes from the inner pot.
	
	Our consideration of this cooling system involves many simplifications. We explore only one particular case of an evaporative cooling system. More general approaches are presented in \cite{boris_halasz},\cite{wu_huang_zhang}.

	\section{Methods}
	
	\subsection{Static equilibrium temperature of the jug}

		\begin{figure}
			\includegraphics[width=0.35\textwidth]{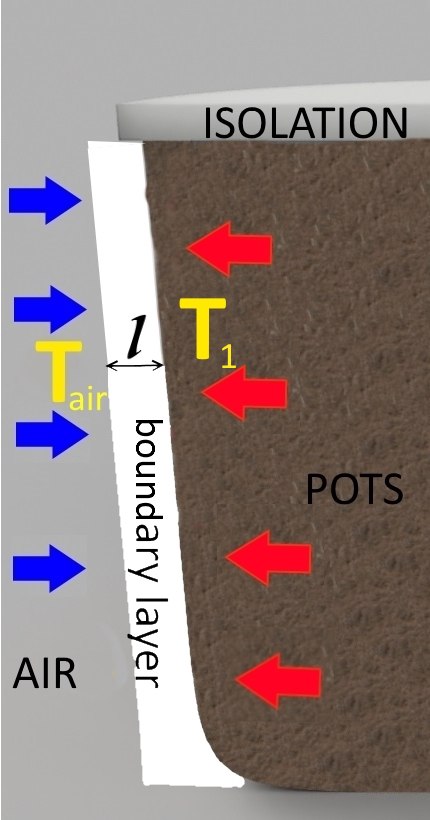}
			\caption{Geometry of the wall of pot-in-pot with boundary layer and the distribution of the heat fluxes at static equilibrium. }
			\label{fig:static_balance}
		\end{figure}

	We will find equilibrium temperature $T_1$ of a wet jug when the air is of temperature $T_\mathrm{air}$ and relative humidity $\phi_\mathrm{air}$. Our model is shown in fig. \ref{fig:static_balance}.
	When $T_\mathrm{air}>T_1$, the surface of the jug  gets heat via heat conduction from the exterior air,
	and uses the heat to evaporate water from the surface.
	The water vapour is then transported away from the surface by diffusion. 
	In equilibrium, the mass of water evaporated due to the heat coming in by heat conduction,
	equals mass of water vapour transported away from the surface by diffusion,
	and it is the equilibrium between heat conduction and diffusion that determines the equilibrium surface temperature $T_1$.
	We assume that both heat and diffusive fluxes propagate across a flat boundary layer of thickness $l$,
	in which gradients of both temperature and absolute humidity are uniform.
	Beyond this layer (at the distance $l$ from the surface) air is assumed to be well mixed,
	and temperature and humidity are assumed to be the same as in the bulk of air.
	We also assume that the jug itself is in thermal equilibrium with its surface, so that the surface exchanges heat only with air,
	and that water covers all the surface, 
	so that the pressure of water vapour at the surface is the same as the pressure of saturated vapour at temperature $T_1$. We compare air temperature at the distances equal or bigger than boundary layer to the temperature of the dry thermometer (air is well mixed there), while temperature of saturated water vapour at the surface of the jug is compared to the wet thermometer.
	Later we are going to analyze these assumptions more scrupulously, and to go beyond them.
	
	Following Newton's law \cite{sivukhin}, the heat flux to the surface is
	\begin{equation}
		J=\kappa \frac{T_\mathrm{air}-T_1}{l}\,,
	\end{equation}
	where $\kappa$ is the heat conductivity of air.
	This heat evaporates the following mass of water from unit surface per unit time,
	\begin{equation}
		\frac{dm}{dt\,dS}=\frac{J}{L}=\frac{\kappa(T_\mathrm{air}-T_1)}{Ll}\,,
		\label{heat_conductivity}
	\end{equation}
	where $L$ is the latent heat of vapourization of water.
	
	The rate at which the evaporated water diffuses away from the surface, is governed by Fick's law,
	\begin{equation}
		\frac{dm}{dt\,dS}=D\frac{\rho(T_1)-\phi_\mathrm{air}\rho(T_\mathrm{air})}{l}\,,
		\label{diffusion}
	\end{equation}
	where $D$ is the diffusion coefficient of water vapour in air, and $\rho(T_1)$ is density of the saturated water vapour as function of temperature.
	
	In equilibrium, eq. (\ref{heat_conductivity}) equals to eq. (\ref{diffusion}), resulting into the following equation for $T_1$,
	\begin{equation}
		\kappa(T_\mathrm{air}-T_1)=DL(\rho(T_1)-\phi_\mathrm{air}\rho(T_\mathrm{air}))\,.
		\label{equaion_for_T}
	\end{equation}
	Importantly, $l$ cancels from this equation, 
	and we do not have to make any assumptions about the boundary layer to determine the equilibrium temperature.
	
\begin{figure}
		\includegraphics[width=0.45\textwidth]{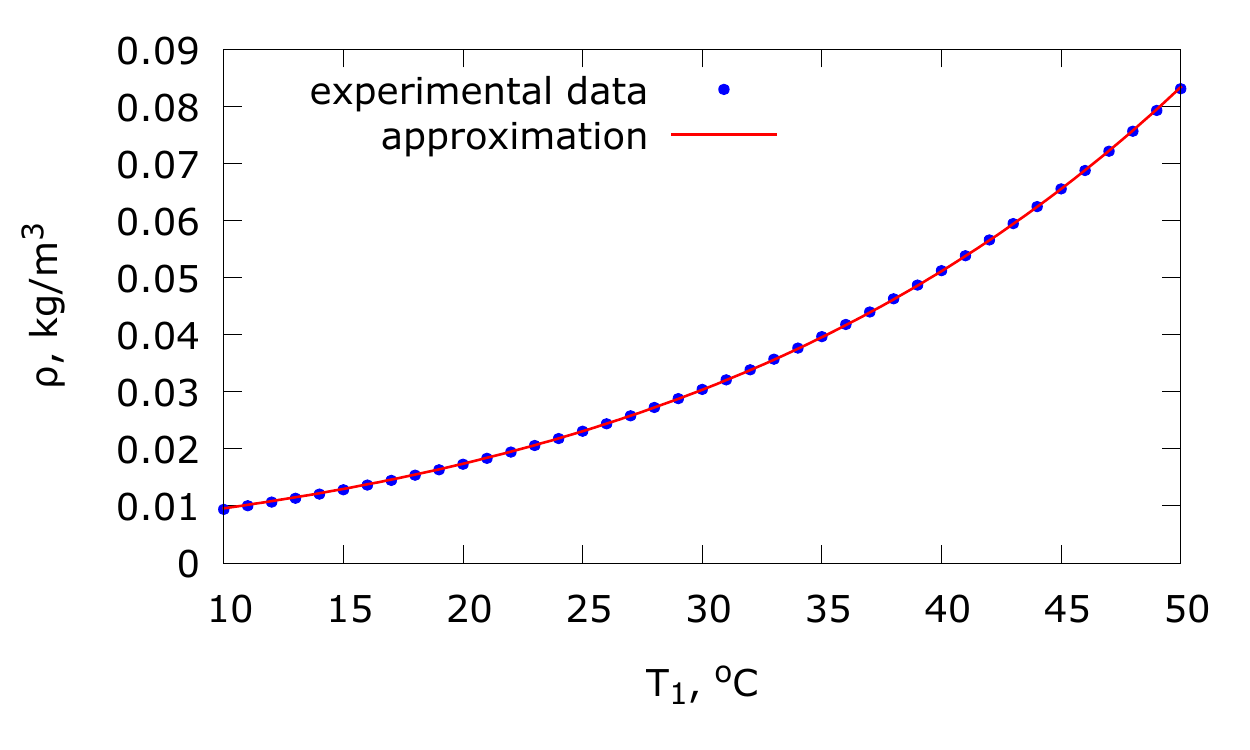}
		\caption{Experimental relation between the temperature and the density of the saturated water vapour \cite{steam_tables}, approximated by eq. (\ref{approximation}).}
		\label{fig:approximation}
	\end{figure}

	To solve this equation, we need to substitute the functional form of $\rho(T_1)$.
	Although such approximations for the pressure dependence $p(T_1)$ are well developed \cite{p(T)1,p(T)2},
	approximations of $\rho(T_1)$ are scarce. For this reason, we had to construct our own approximation.
	We chose a functional form similar to the August equation or, more generally, to Clausius-–Clapeyron equation,
	\begin{equation}
		\rho(T_1)=A\exp\left(-\frac{B}{T_1}\right)\,,
		\label{approximation}
	\end{equation}
	where the temperature $T_1$ is given in degrees Kelvin, while $A$ and $B$ are free constants, 
	chosen to best fit the experimental data in the temperature range $10-50^\circ\mathrm{C}$. In the Fig.(\ref{approximation}) 50 $\circ \mathrm{C}$ is the theoretical extrapolation, our approach can not define humidity precisely at such temperatures.
	The best fit is provided by $A=369000$ kg m$^{-3}$, $B=4943$ K.
	It is shown in fig. \ref{fig:approximation}, overplotted with the experimental data \cite{steam_tables}.
	
	\begin{figure}
		\includegraphics[width=0.45\textwidth]{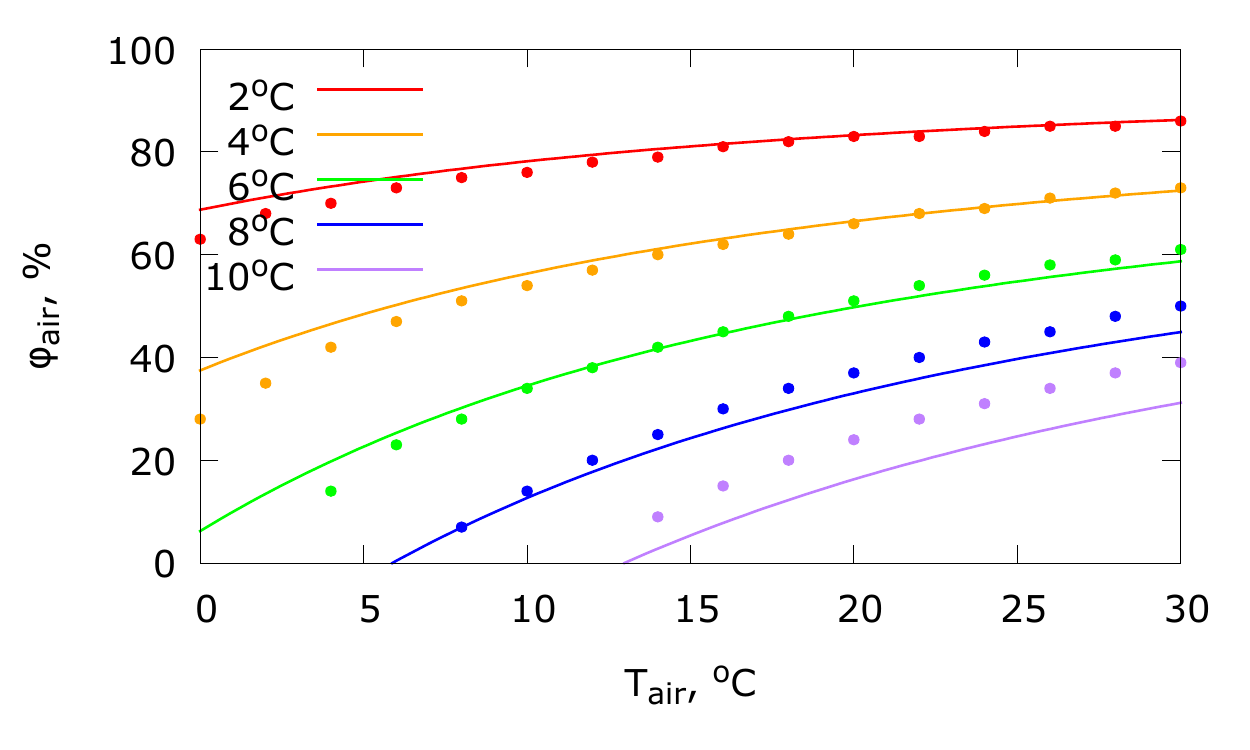}
		\caption{Data from the psychrometric table (circles) overplotted with the theoretical formula - eq. (\ref{phi}) (solid lines).
			The table shows humidity $\phi_\mathrm{air}$ versus temperature $T_\mathrm{air}$ of the exterior air (dry thermometer).
			Different temperature differences $\Delta T$ between the dry and the wet thermometer are plotted in different colours.}
		\label{fig:humidity}
	\end{figure}
	
	\begin{figure}
		\includegraphics[width=0.45\textwidth]{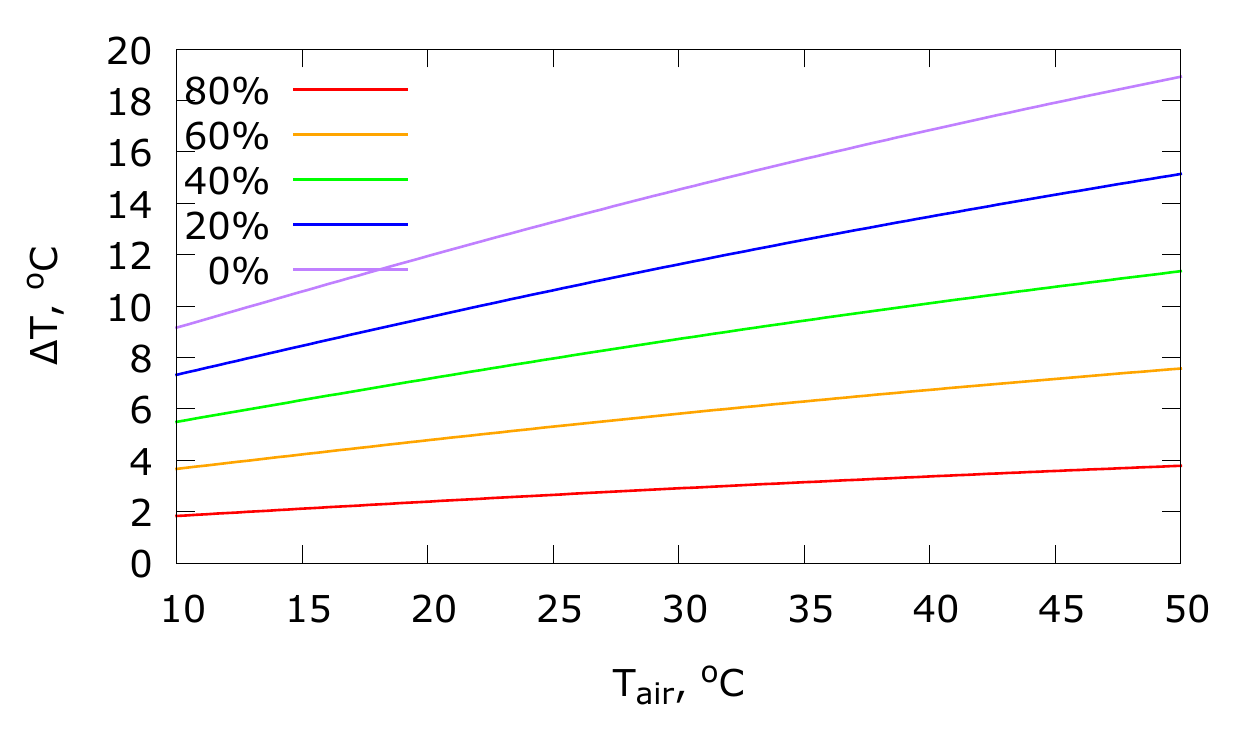}
		\caption{
			The temperature difference $\Delta T$ between the cooling jug and the exterior air as given by eq. (\ref{dT1}),
			plotted against the temperature of the exterior air $T_\mathrm{air}$.
			Situations corresponding to different humidities $\phi_\mathrm{air}$ are shown in different colours.}
		\label{fig:temperature}
	\end{figure}
	
	When we substitute eq. (\ref{approximation}) into eq. (\ref{equaion_for_T}), we get a transcendental equation,
	that cannot be solved using elementary functions.
	Still, we notice that $\Delta T=T_\mathrm{air}-T_1$ is usually only a few degrees Celsius, 
	so that the difference between $\rho(T_\mathrm{air})$ and $\rho(T_1)$ is small.
	That allows us to decompose $\rho(T_1)=\rho(T_\mathrm{air}+\Delta T)$ into a Taylor series in terms of $\Delta T$, 
	truncating the series after the linear term.
	The general form of the solution to the resulting linearized equation is
	\begin{equation}
		\Delta T=\frac{(1-\phi_\mathrm{air})\rho(T_\mathrm{air})}{\rho'(T_\mathrm{air})+\frac{\kappa}{DL}}\,.
		\label{dT1}
	\end{equation}
	\\
	To verify our method, we reverse eq. (\ref{dT1}) using eq. (\ref{approximation}) to obtain humidity,
	\begin{equation}
		\phi_\mathrm{air}=1-\Delta T\bigg(\frac{B}{T_\mathrm{air}^2}+\frac{\kappa}{DLA}\exp\left(\frac{B}{T_\mathrm{air}}\right)\bigg)\,.
		\label{phi}
	\end{equation}
	Then in fig. \ref{fig:humidity} we overplot this theoretical expression for humidity with data from a psychrometric table\footnotemark \cite{koshkin}.
	\footnotetext{Psychrometric table is used to determine relative humidity based on measurement results of a psycrometer, i.e. a device consisting of two thermometers, a dry one and a wet one. The dry thermometer shows the temperature of the ambient air, while the thermometer wrapped in wet fabric shows the temperature of evaporating water. The readings of the thermometers can then be translated into the relative humidity of the air.}
			
	We see a decent agreement between the data and the theoretical prediction, which justifies the methods we are using.
	Still, for big temperature differences $\Delta T$ we see an increasing discrepancy between the theory and the data.
	This should be expected, as the accuracy of the Taylor approximation decreases with $\Delta T$.

	Next we use eq. (\ref{dT1}) to predict the $\Delta T$ for the cooling jug, and plot the results in fig. \ref{fig:temperature}.
	As expected, the cooling $\Delta T$ is larger when the ambient temperature is high, and the humidity low.
	According to our theory, at $T_\mathrm{air}=50^\circ\mathrm{C}$ the temperature difference can reach up to $20^\circ\mathrm{C}$, although as we see from fig. \ref{fig:humidity} the theory tends to give bigger errors at big $\Delta T$.
	
	Although our approach is less precise than the available phenomenological formulae \cite{precise_temp}, it gives a clearer understanding of the underlying physics, and allows for simple generalizations to non-stationary cases, as we demonstrate in the following  subsection.

	\subsection{Dynamics of jug cooling}
\begin{figure}
		\includegraphics[width=0.45\textwidth]{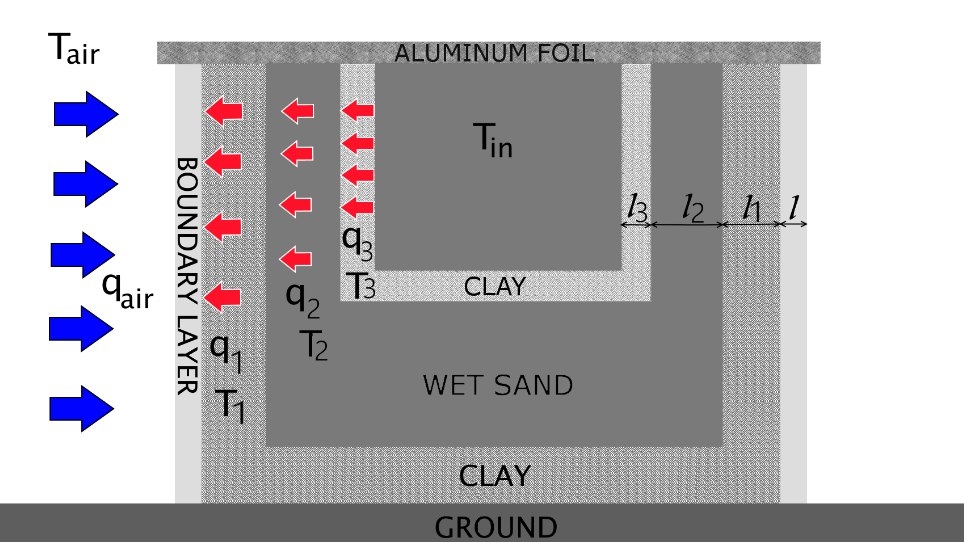}
		\caption{Geometry of  pot-in-pot cooler and the distribution of the heat fluxes at dynamic equilibrium}
		\label{fig:flux_distr}
	\end{figure}

	We want to consider how the finite heat conductivity of the jug walls hinder its cooling speed. We assume the jug to be filled with a well mixed liquid, whose temperature $T_\mathrm{in}$ is uniform throughout its volume, and whose specific heat capacity is $c_\mathrm{liq}$. The wall consists of three layers, as shown in Fig. \ref{fig:flux_distr}. Two layers are composed of clay and one is composed of sand. 

	To focus on the evaporation itself, we consider a simplified geometry where the wall thickness of both jugs, the distance between the jugs and the thickness of the air boundary layer are all much smaller than the inner radius of the inner jug. This allows us to treat the heat conduction and diffusion problems as one-dimensional. The theory required to go beyond this assumption will be briefly reviewed in Discussion.
	
	So far we have only considered the influence of the outside heat flux on water evaporation from the surface. The inner heat flux should be also taken into account. The heat fluxes in the different layers $q_1, q_2, q_3$ and $q_\mathrm{air}$ (see Fig. \ref{fig:flux_distr}) are determined by the heat conductivity equation:
	\begin{equation}
	q_\mathrm{air} = \kappa_\mathrm{air}\frac{T_\mathrm{air}-T_1}{l}\,,
	\label{q_air}
	\end{equation}
	\begin{equation}
	q_1 = \kappa_\mathrm{clay}\frac{T_2-T_1}{l_1}\,,
	\label{q_1}
	\end{equation}
	\begin{equation}
	q_2 = \kappa_\mathrm{sand}\frac{T_3-T_2}{l_2}\,,
	\label{q_2}
	\end{equation}
	\begin{equation}
	q_3 = \kappa_\mathrm{clay}\frac{T_\mathrm{in}-T_3}{l_3}\,.
	\label{q_3}
	\end{equation}

	The heat flow from inside the jug $q_1$ contributes to water evaporation from the surface. The sum of the outer and the inner heat flows to the surface is the reason for water evaporation, as it is represented by the heat balance equation,
	\begin{equation}
	q_\mathrm{air}+q_1=L\frac{dm}{dt\,dS} \, .
	\label{surface_balance}
	\end{equation}

	The same heat flow $q_1$ causes the jug to cool down in accordance with the energy balance equation,
	\begin{equation}
	C\frac{dT_\mathrm{in}}{dt}=-q_1 S \, .
	\label{interior_balance}
	\end{equation}
	Here $S$ is again the total surface area of the jug, and $C$ is the total heat capacity of the jug,
	\begin{equation}
	C = m_\mathrm{liq}c_\mathrm{liq} + m_\mathrm{sand}c_\mathrm{sand} + m_\mathrm{clay}c_\mathrm{clay}\,,
	\label{heat_capacity}
	\end{equation}
	with $m$ and $c$ with different subscripts denoting masses and specific heat capacities of liquid, sand and clay.

	These equations allow us to calculate cooling rate of the jug. We assume that the thermal inertia of the liquid inside the jug is much bigger than the thermal inertia of the jug itself. Then approximately $q_1=q_2=q_3$, which allows us to eliminate $T_2$ and $T_3$ from eqs. (\ref{q_1})-(\ref{q_3}), and to express $q_1$ in terms of $T_1$ and $T_\mathrm{in}$. Then we substitute this equation together with eq. (\ref{q_air}) into eq. (\ref{surface_balance}), express $\frac{dm}{dt\,dS}$ from eq. (\ref{diffusion}), and thus get an equation for $T_1$. Again applying the Taylor decomposition to $\rho(T)$, we linearize this equation, and easily find its solution $T_1$. From the resulting $T_1$ we calculate $q_1$, and substitute it into eq. (\ref{interior_balance}). Thus we arrive at a linear differential equation for $T_\mathrm{in}$, whose solution is
	\begin{equation}
	T_\mathrm{in}= T_\mathrm{air}-\Delta T+\left(T_\mathrm{liq0}-T_\mathrm{air}+\Delta T\right)\mathrm{e}^{-\frac{t}{\tau}}\, ,
	\label{T_final}
	\end{equation}
	with the relaxation time $\tau$ determined as
	\begin{equation}
	\tau=\frac{C}{S}\left(\frac{l_1+l_3}{\kappa_\mathrm{clay}}+\frac{l_2}{\kappa_\mathrm{sand}}+\frac{l}{\kappa_\mathrm{air}+LD\rho'}\right) \, .
	\label{tau_final}
	\end{equation}
	
	\begin{table}
		\caption{Values of the parameters used in our models}
		\begin{tabular}{l l l l}
			Symbol & Value \\
			\hline	
			$L$ & $(2.46\pm 0.02) \cdot 10^{6}$ $\mathrm{J \cdot kg^{-1}}$ \\
			$D$ & $(2.1\pm 0.1) \cdot 10^{-5}$  $\mathrm{m^2 \cdot s^{-1}}$ \\
			$c_{\mathrm{liq}}$ & $(4.20\pm 0.02)\cdot 10^{3}$ $\mathrm{J} \cdot \mathrm{kg}^{-1} \cdot \mathrm{K}^{-1}$ \\
			$c_{\mathrm{clay}}$ & $(9.0\pm 0.2)\cdot 10^{2}$ $\mathrm{J} \cdot \mathrm{kg}^{-1} \cdot \mathrm{K}^{-1}$ \\
			$c_{\mathrm{sand}}$ & $(8.2\pm 0.2)\cdot 10^{2}$ $\mathrm{J} \cdot \mathrm{kg}^{-1} \cdot \mathrm{K}^{-1}$ \\
			$\kappa_{\mathrm{clay}}$ & $(1.5\pm 1)$  $\mathrm{W \cdot m^{-1} \cdot K^{-1}}$ \\
			$\kappa_{\mathrm{sand}}$ & $(2.5\pm 0.5)$  $\mathrm{W \cdot m^{-1} \cdot K^{-1}}$ \\	
			$\kappa_{\mathrm{air}}$ & $(2.57\pm 0.04) \cdot 10^{2}$  $\mathrm{W \cdot m^{-1} \cdot K^{-1}}$ \\	
			$v$ & $(1\pm 0.7)$ $\mathrm{m \cdot s^{-1}}$ \\
			$\eta$ & $(1.72\pm 0.03) \cdot 10^{-5}$ $\mathrm{Pa \cdot s}$ \\
			$\alpha$ & $(5\pm 2)$	
		\end{tabular}
		\label{tab1}
	\end{table}
	
	The boundary layer for a laminar flow can be expressed in the following way \cite{boundary_layer_ref}:
	\begin{equation}	
	l=\alpha\sqrt{\frac{\eta R}{\rho v}}\,.
	\label{boundary_layer}
	\end{equation}
	Here $\eta$ is the dynamic viscosity, $R$ is the characteristic length of the pot, $\rho$ is the density of the oncoming flow, and $v$ is the velocity of this flow, and $\alpha$ is a constant of the order of unity, which is commonly set at $\alpha=5$ \cite{boundary_layer_ref}. Then the thickness $l$ corresponds to the distance from the surface, where the air speed reaches 99\% of the speed of the wind \cite{boundary_layer_ref}. Assumption that the air is well mixed just outside this area and poorly mixed inside is somewhat arbitrary, and setting different speed limits instead of 99\% would correspond to a different $\alpha$. Moreover, the boundary layer has different thickness at different parts of the jug, thus making eq. (\ref{boundary_layer}) a mere estimate. To account for these uncertainties, we take $\alpha=5\pm 2$.

	Other parameters relevant for our model are listed in Table \ref{tab1} together with their uncertainties.
$L$, $D$, $c_{\mathrm{liq}}$, $\kappa_{\mathrm{air}}$ and $\eta$ are given at room temperature, and the uncertainty is computed by assuming temperature variations of 10$^\circ$C. Uncertainties of $c_{\mathrm{clay}}$, $c_{\mathrm{sand}}$, $\kappa_{\mathrm{clay}}$ strongly depend on the kind of sand and clay used, and are estimated from the range of values cited for these substances (e.g. \cite{toolbox, clay_conductivity}). Uncertainty in $\kappa_{\mathrm{sand}}$ is predominantly dictated by the uncertain wetness of sand \cite{wet_sand_heat_conductivity}. For the speed of wind we took typical speeds of air in a room.

	Other important limitations of the proposed theory include different surface areas of different boundaries, as well as inequality of heat fluxes at these boundaries, because some heat is stored in heat capacity of the material. Heat exchange through the top and the bottom of the jug was also neglected.

	It is interesting though, that despite these shortcomings, our model can give reasonable estimates for the cooling rate, as we will see in the following section.

	\section{Results}
	
	\subsection{Experiment}
	For the experiment we use the two clay pots of different size. The space between the pots is filled with sand. Although we give theory for the case when the inner pot is filled with water, our experiment is carried out with wet sand. Such replacement is appropriate as we only need to change specific heat capacity and heat conductivity in our calculations. In order for refrigeration to occur, we need moisture, so we pour water into the sand between the pots. The sand becomes wet. Water enters the micropores of unbaked clay, due to the capillary effect. Shortly after the start of the experiment, we observe the formation of droplets on the outside of the pot. Drops evaporate from the entire surface of the pot and the system is cooled.

	If the outer pot is made of baked clay or covered with paint, the micropores on the surface are closed, and water is not able to appear on the outer surface for evaporation.

	In order to neglect the evaporation of water directly from the space between pots and to reduce heat loss to the environment the system is sealed with foil on top.

	The humidity of the ambient air is $\phi_\mathrm{air} = 55\%$. The thermometer is put inside of the cup. The room temperature is $18.0^\circ\mathrm{C}$. After 5 hours the system temperature is equal to $13.5^\circ\mathrm{C}$. Further, the temperature slowly decreases, approximately following an exponential law. In order for the system to continue cooling, we need to add water, because otherwise the sand becomes drier. In the beginning of the experiment we poor about 0.5 liters of water into the sand. Then we add water every half an hour, resulting in the total of 300 ml of water added. As the heat capacity of the added water is much smaller than the heat capacity of sand, adding the water could not much influence the cooling rate of the system, although certainly should have been accounted for in a more precise theory.

\begin{figure}
				\includegraphics[width=0.4\textwidth]{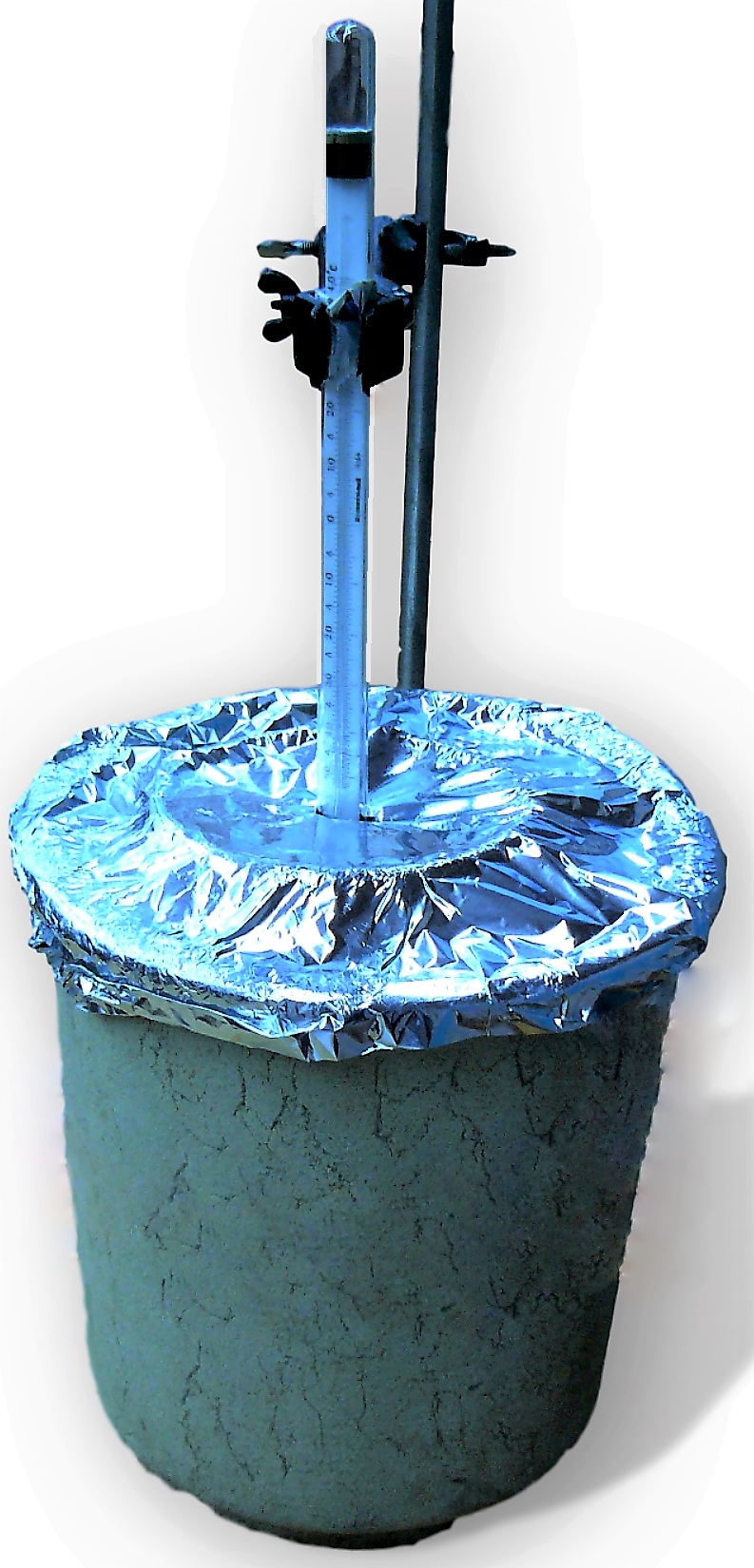}
				\caption{View of the experimental system. Its properties are as follows. Inner pot: Height - $15\mathrm{cm}$, Diameter - $9\mathrm{cm}$, Thickness - $6\mathrm{mm}$.
					Outer pot: Height - $40\mathrm{cm}$, Diameter - $18\mathrm{cm}$, Thickness - $6\mathrm{mm}$.
					We used 10 kg of sand and 0.8 litres of water.}
				\label{fig:setup}
			\end{figure}

	Our setup is shown in fig. \ref{fig:setup}. Noteworthy, in our experiment the wall thickness of the jug is not small as compared to the radius, in direct contradiction with assumptions of the theory. Therefore, our theory can describe the experiment only approximately.
	
\begin{figure}
				\includegraphics[width=0.45\textwidth]{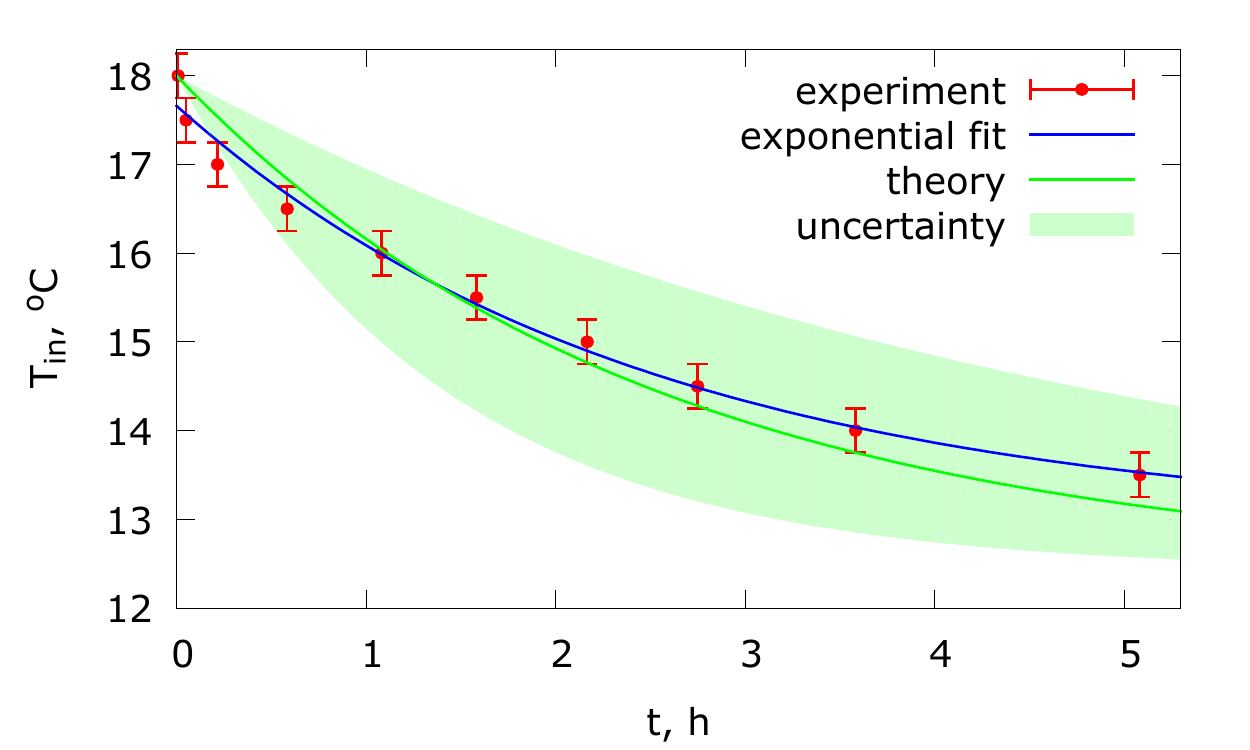}
				\caption{Results of the experiment (red dots) overplotted with the least-square exponential fit (blue line) and the theoretical predictions by eq. (\ref{T_final}) (green line). The area shadowed in light green shows the theoretical uncertainty.}
				\label{fig:experiment}
			\end{figure}

	Temperature versus time is shown in fig. \ref{fig:experiment}. Substitution of the relevant physical parameters and their uncertainties into eq. (\ref{T_final}) with $\Delta T=5.6^\circ$C and $\tau=2.5_{-1.1}^{+2.3}$ hours. This theoretical uncertainty is shown in the figure as light-green shaded area.
	
	Note, that one experiment is not sufficient to prove a theory with so many parameters, but merely to illustrate it. Moreover, giving substantial uncertainties in the parameters involved in the theory (especially the coefficient in eq. (\ref{boundary_layer})), such a good agreement of the theory and the experiment should be coincidental to some extent. Still, it allows us to argue, that our theory well reproduces the general temperature trend and gives at least a reasonable order-of-magnitude estimate for the cooling rate.
	
	\section{Discussion} 
		
	Some physical factors omitted by our theory could in principle be incorporated into the model,
	although their inclusion would make the model overly complicated analytically, while their practical significance is hard to estimate.
	Here we only briefly discuss the underlying physics, and its effect on our model.
	
	\subsection{Geometric shape}
	\label{sec:shape}
	Our previous calculations were done under the assumption of a jug with a thin shell and an interior of a constant temperature.
	To understand the cooling rate of the shell or the temperature inhomogeneity in the interior, we must do a rigorous 3D simulation of heat diffusion in the jug.
	This means going from an ordinary differential equation like eq. (\ref{interior_balance}) to a partial differential equation.
	
	Matter and energy balance on the surface results into the boundary condition for this PDE. To derive it, we equate equations (\ref{heat_conductivity}) and (\ref{diffusion}). By transforming and simplifying this expression, we get an inhomogeneous boundary condition of the third kind:
	
	\begin{equation}
	\alpha \left. \frac{\partial T}{\partial r} \right|_{r=R}+ \beta \left. T \right|_{r=R} = \gamma \,,
	\label{third_kind_boundary_condition}
	\end{equation}
with the coeffients
	\begin{equation}
	\alpha = l \kappa _{clay}\,,
	\label{alpha}
	\end{equation}
	
	\begin{equation}
	\beta = -\left( \kappa_{air} + LD \rho ' (T_0) \right)\,,
	\label{beta}
	\end{equation}
	
	\begin{eqnarray}
	\gamma &=& LD\left( (1-\phi _0) \rho(T_0) - \rho ' (T_0) T_0 \right) -\nonumber\\
	&-& \kappa_{air} T_0\,.
	\label{gamma}
	\end{eqnarray}
	
	To get the temperature distribution, the heat conductivity equation has to be solved:
	\begin{equation}
	\frac{\partial T}{\partial t} = a^2 \Delta T\,.
	\label{heat_conductivity_equation}
	\end{equation}	
	Here $\Delta$ denotes the Laplace operator.
	
	For a cylindrical pot, the Laplace operator can be rewritten in cylindrical coordinates,
	and the solution will be obtained as a series in terms of the Bessel functions,
	although the coefficients of this series can be found only from a transcendental algebraic equation.
	For a spherical pot, the problem can be simplified in spherical coordinates,
	and the solution can be obtained in a similar manner as a series in terms of spherical functions.
	For a pot of an arbitrary shape, there are no general methods to solve the equation,
	and it should be simulated numerically.
	
	\subsection{Viscosity and surface tension}
One more limitation to water evaporation from the surface is set by the viscosity of water. On its way through the clay, water has to overcome viscous friction in capillaries, which in the case of a cylindric capillary is governed by Hagen--Poiseuille equation,
	\begin{equation}
	\frac{dm}{dt}=\frac{8 \mu L}{\pi\rho R^4}\Delta p\,.
	\label{poisseil}
	\end{equation}
Here $\mu$ is the dynamic viscosity of water, $\rho$ is its density, $L$ is the length of the capillary, $R$ is its radius, $\Delta p$ is the pressure difference between the two ends of the capillary, and $\frac{dm}{dt}$ is the mass flow rate of water.

If the sand is not soaking wet and can hold water inside, water pressure inside it is not greater than the atmospheric pressure. Then to move water through the capillary, the water pressure on the outer side of the jug must be less than the atmospheric pressure by some amount $\Delta p$, which could only be attributed to the Laplace pressure on the outer side of the capillary,
	\begin{equation}
	\Delta p = \frac{2\sigma}{R}\,.
	\label{laplace}
	\end{equation}
Here $\sigma$ is water surface tension, and the radius of curvature $R$ is of the same order of magnitude as the radius of the capillary.

Now we can see the non-trivial interplay between these phenomena. If water experiences high friction in capillaries, the water pressure on the outer surface of the jug falls. Now water is sucked inside the thinnest capillaries, which have the biggest Laplace pressure. But the same Laplace pressure makes it harder for water to evaporate, and diminishes the pressure of saturated vapour on the surface of the jug. By this mechanism viscosity decreases the cooling rate of the jug.
	
\section{Dead-ends}
			\begin{figure}
				\includegraphics[width=0.45\textwidth]{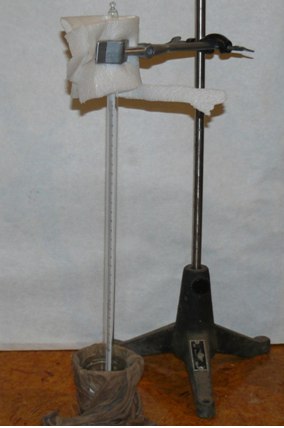}
				\caption{The experimental setup used for our fist experiments.}
				\label{fig:old_pot}
			\end{figure}
			\begin{figure}
				\includegraphics[width=0.45\textwidth]{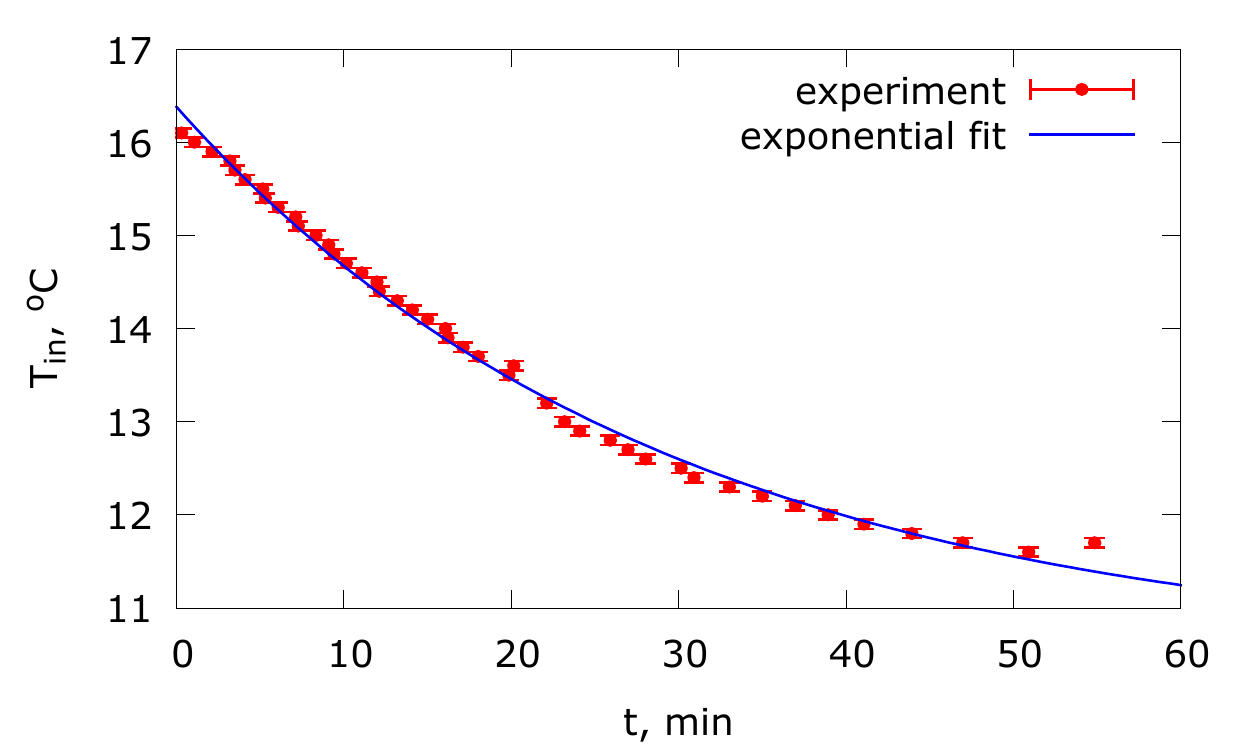}
				\caption{Results of our first experiment (red dots) ant the least-square exponential fit (blue line).}
				\label{fig:old_temperature}
			\end{figure}
The experiments, with which we started modelling this problem and which we presented at the tournament, were much cruder than the experiments described above. The major problem was that we did not have pots made of raw clay. So we used materials at hand to construct a setup shown in fig. \ref{fig:old_pot}.

A paper cup was covered with a cloth rag from inside and outside. Another cup was placed inside it. Water was put in the first cup, as well as to the space between the cups.  Water rose by the cloth due to the capillary effect, and evaporated, cooling the setup. The thermometer was put inside the cup. The room temperature was 16.2$^\circ$C, and in less than an hour the water temperature inside the cup fell down to 11.5$^\circ$C. Fig. \ref{fig:old_temperature} shows the temperature as a function of time. The general trend is similar to fig. \ref{fig:experiment}, yet the exponential law now provides a worse fit to the data.

Although representing the same physics, this model significantly deviated from the problem statement. Moreover, water oozing through a folded cloth of complicated geometry presents a problem, whose precise physical description is not easier than the original problem of pot-in-pot refrigerator.

	\section{Conclusions}
	The cooling of the pot is caused by water evaporation from its surface.
	The minimal temperature attainable by the pot is determined from the mass and energy conservation laws at the surface.
	The rate at which this temperature will be achieved is limited by the heat conductivity through the pot, the heat conductivity through the boundary layer, and the diffusion of water vapour through the boundary layer.
	
	In the simplest model we get a linearized algebraic equation for the minimal temperature, and a linear differential equation for the process of cooling to this temperature. The solution of the former is in reasonably good agreement with the psychrometric table. The solution of the latter as a function of time is an exponential function, which qualitatively agrees with our experiments.
	
	To make the model more precise, one can include additional factors, such as the complex geometric shape of the jug, the influence of resistance to water flow in the capillaries, the curvature of the water surface in the capillaries, and air convection as a mixing agent that supplements the exterior wind. But this added precision will come on an intolerable cost - tremendously increased complexity.

\end{document}